\documentclass[9pt]{article}
\usepackage{spconf,amsmath,graphicx,xcolor,float}
\usepackage{booktabs,hyperref,subcaption}

\title{C3DVQA: Full-Reference Video Quality Assessment with 3D Convolutional Neural Network}
\address{{$^1$}School of Electronic and Computer Engineering, Peking University Shenzhen Graduate School\\
	 {$^2$}Media Lab, Tencent ~~~ {$^3$}Peng Cheng Laboratory }
\begin{document}
\makeatletter
\def\@name{\textit{Munan Xu{$^{1,3}$}, Junming Chen{$^{1,3}$}, Haiqiang Wang{$^{2}$}, Shan Liu{$^{2}$}, Ge Li{$^{1,3}$}, Zhiqiang Bai{$^{1}$}}\\}
\makeatother
\maketitle

\begin{abstract}
Traditional video quality assessment (VQA) methods evaluate localized picture quality and video score is predicted by temporally aggregating frame scores. However, video quality exhibits different characteristics from static image quality due to the existence of temporal masking effects. In this paper, we present a novel architecture, namely C3DVQA, that uses Convolutional Neural Network with 3D kernels (C3D) for full-reference VQA task. C3DVQA combines feature learning and score pooling into one spatiotemporal feature learning process. We use 2D convolutional layers to extract spatial features and 3D convolutional layers to learn spatiotemporal features. We empirically found that 3D convolutional layers are capable to capture temporal masking effects of videos. We evaluated the proposed method on the LIVE and CSIQ datasets. The experimental results demonstrate that the proposed method achieves the state-of-the-art performance.
\end{abstract}


\section{Introduction}
\label{sec:intro}

%%%%%%%%%%%%%%%%%%%%%%%%%%%%%%%%%%%%%%%%%%%%%%%%%%%%%%%%%%%%%%%%%%
\begin{figure*}[!ht]
\centering
	\begin{subfigure}[b]{1.0\linewidth}
	\centering{}
	\includegraphics[height=0.25\linewidth]{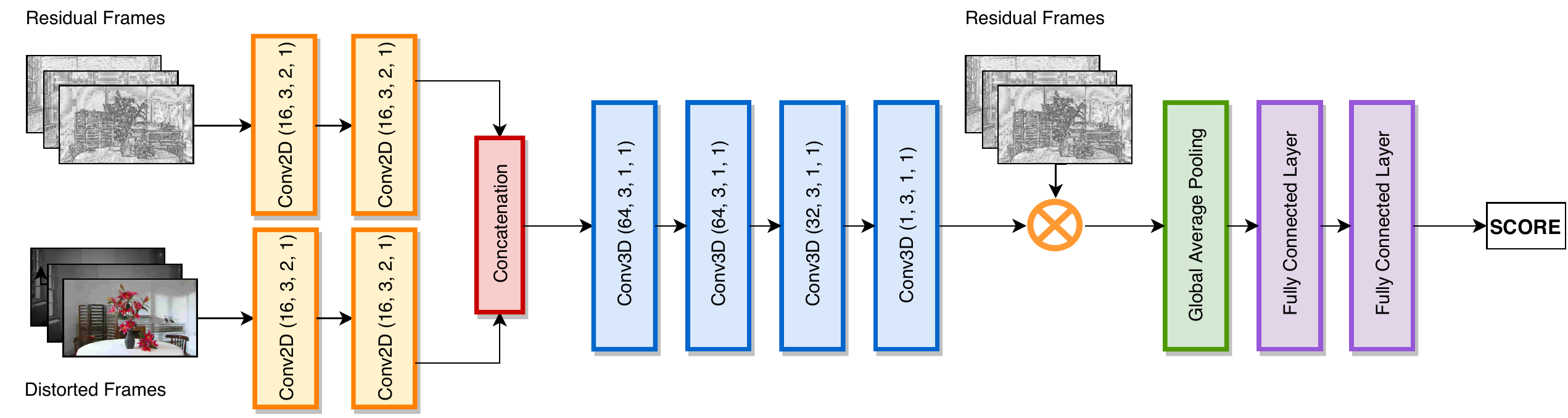}
	% \caption*{\label{}}
	\end{subfigure}
\caption{Architecture of the proposed C3DVQA network. It consists of 2D convolutional layers to learn spatial features, 3D convolutional layers to learn spatiotemporal features and regression layers. Elements in the tuple of convolutional layers indicate the number of channels, kernel size, stride and padding, respectively.} \label{fig:C3D_architecture}
\end{figure*}
%%%%%%%%%%%%%%%%%%%%%%%%%%%%%%%%%%%%%%%%%%%%%%%%%%%%%%%%%%%%%%%%%%

Real-time video streaming services consume a large amount of Internet traffic.
It is desired to reduce bandwidth requirement without degrading viewing experience. Thus, video quality assessment (VQA) technology has attracted a lot of attention. There have been several image quality assessment (IQA) metrics with Convolutional Neural Network (CNN) proposed in the literature \cite{bosse2017deep,kim2017deep,liu2017rankiqa}. The predicted quality score correlates well with the subjective score of static images. However, it is still a challenging task to develop quality metrics that can accurately measure the perceived quality of video content.

An important characteristic to consider about video quality is the temporal masking effect. The masking effect refers to the reduced capability of the human visual system (HVS) to detect a stimulus with a spatially or temporally complex background. It is generally accepted that image quality is primarily determined by distortions masked by the spatial masking effect \cite{daly1992visible}. However, video quality is further influenced by the temporal masking effect. The distortion threshold is jointly adjusted by the spatial and temporal masking effect. Besides, motion-related distortions also have an impact on the perceived quality. Thus, it is essential to simultaneously exploit spatial and temporal characteristics to develop a video quality metric.

In order to incorporate motion information in VQA metrics, an intuitive strategy is to apply IQA metrics on spatiotemporal slices of videos. The score of the entire video is obtained by pooling individual slice scores. The slices could be along the spatial axes or the spatiotemporal axes \cite{vu2011spatiotemporal,vu2014vis3}. It was reported \cite{vu2014vis3} that slices along the time axis could characterize changes of motion over time. Recently, several studies conducted analysis on 3D spatiotemporal segments in the transformed domain. Li \emph{et.al.}in \cite{li2016spatiotemporal} adopted 3D discrete cosine transform (3D-DCT) and the approach in \cite{li2015no} employed 3D shearlet transform. Spatiotemporal coefficients or statistics were used to evaluate video integrity.

Inspired by the breakthrough of deep learning approaches in image tasks, 2D CNN methods were extended to CNN with 3D convolutional kernels \cite{ji20123d} for video-related tasks. A spatiotemporal feature learning approach was proposed \cite{tran2015learning} and it was shown that the learned features with a simple linear classifier could yield good performance on various video tasks. 3D convolutional layers were used in \cite{liu2018end,zhang2018deepqoe} to learn spatiotemporal features that are essential to the VQA task.

In this paper, we propose a spatiotemporal feature learning framework using 3D CNN for the full-reference video quality assessment(FR-VQA) task. We apply 2D convolutional layers on the distorted and residual frames to learn spatial features. The learned features are concatenated together to represent the spatiotemporal context of videos. The concatenation is followed by 3D convolutional layers to learn the spatiotemporal distortion threshold. Noticeable artifacts are obtained by masking residual frames with the corresponding distortion threshold. Finally, we use fully connected layers to learn the nonlinear relationship between masked artifacts and the subjective score.
%Experimental results are given to demonstrate the performance of the proposed architecture.

The rest of the paper is organized as follows. The proposed C3DVQA architecture is detailed in Section \ref{sec:method}. Experimental results are presented in Section \ref{sec:results}. Conclusion remarks are given in Section \ref{sec:conclusion}.

\section{C3D network to learn distortion visibility threshold}
\label{sec:method}

In this section, we present the proposed architecture to learn the distortion visibility threshold of videos using CNN with 2D and 3D convolutional kernels. We also introduced the nonlinear regression between masked distortions and subjective quality of videos.

The proposed C3DVQA architecture is given in Fig. \ref{fig:C3D_architecture}. The inputs are the luminance channel of the distorted and residual frames. The residual frames are the differences between reference and distorted frames. We use two 2D convolutional layers and four 3D convolutional layers in the network, followed by one global average pooling layer and two fully connected layers.

\subsection{Convolutional Layers}
\label{3D-CNN}

It is common to treat a video as static images and apply pre-trained 2D CNN model on images for video-related tasks. However, this strategy does not perform well for motion sensitive tasks because the motion information is simply ignored \cite{tran2015learning}. Since video quality is highly correlated with the degree of motion between consecutive frames, we apply 3D CNN to jointly learn spatiotemporal features.

Let $\mathbf{x}$ denote the input spatiotemporal segment of size $C \times D \times H \times W$, where $C$ is the number of channels, $D$ is the number of frames, $H$ and $W$ are the patch height and width, respectively. 2D CNN layers apply 3-dimensional filters on localized patches. Each layer has size $N_{i} \times d \times d$, where $N_{i}$ and $d$ denote the number and the spatial extension of filters, respectively. 3D CNN layers preserve temporal information by applying 4-dimensional tensors on localized spatiotemporal segments. Each layer has size $N_{i} \times t \times d \times d$, where $t$ denotes temporal extension of filters.

In the proposed architecture, the size of 2D convolutional layers is $16 \times 3 \times 3$. We use spatial padding 1 and stride 2 to spatially downsample the input with each convolutional layer.
After two 2D convolutional layers, the dimension of the output is $16 \times D \times H/4 \times W/4$.
The outputs of residual and distorted features are concatenated together to represent spatiotemporal context that are essential to the video quality. The 3D convolutional layers have the kernel size of $3 \times 3 \times 3$ and the number of channels are 64, 64, 32 and 1, respectively. We do not downsample spatially or temporally by applying spatiotemporal padding 1 and stride 1. The output of 3D convolutional layers has a size of $1 \times D \times H/4 \times W/4$ and represents the spatiotemporal distortion visibility threshold.

It should be noted that the convolutional layers are designed to keep the temporal dimension unchanged, i.e. $d=D$. The purpose is to apply a frame by frame threshold masking on residual frames. In this way, the network is able to capture localized artifacts on random frames.

%%%%%%%%%%%%%%%%%%%%%%%%%%%%%%%%%%%%%%%%%%%%%%%%%%%%%%%%%%%%%%%%%%
\begin{figure*}[!htb]
\centering
	\begin{subfigure}[b]{0.12\linewidth}\centering
		\includegraphics[width=\linewidth]{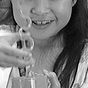}\caption{}\label{res_a}
	\end{subfigure}
	\begin{subfigure}[b]{0.12\linewidth}\centering
		\includegraphics[width=\linewidth]{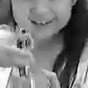}\caption{}\label{res_b}
	\end{subfigure}
	\begin{subfigure}[b]{0.12\linewidth}\centering
		\includegraphics[width=\linewidth]{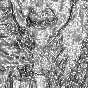}\caption{}\label{res_c}
	\end{subfigure}
	\begin{subfigure}[b]{0.12\linewidth}\centering
		\includegraphics[width=\linewidth]{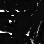}\caption{}\label{res_d}
	\end{subfigure}
	\begin{subfigure}[b]{0.12\linewidth}\centering
		\includegraphics[width=\linewidth]{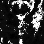}\caption{}\label{res_e}
	\end{subfigure}
	\begin{subfigure}[b]{0.12\linewidth}\centering
		\includegraphics[width=\linewidth]{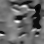}\caption{}\label{res_f}
	\end{subfigure}
	\begin{subfigure}[b]{0.12\linewidth}\centering
		\includegraphics[width=\linewidth]{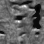}\caption{}\label{res_g}
	\end{subfigure}

	\begin{subfigure}[b]{0.12\linewidth}\centering
		\includegraphics[width=\linewidth]{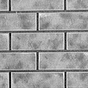}\caption{}\label{res_h}
	\end{subfigure}
	\begin{subfigure}[b]{0.12\linewidth}\centering
		\includegraphics[width=\linewidth]{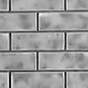}\caption{}\label{res_i}
	\end{subfigure}
	\begin{subfigure}[b]{0.12\linewidth}\centering
		\includegraphics[width=\linewidth]{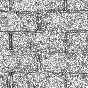}\caption{}\label{res_j}
	\end{subfigure}
	\begin{subfigure}[b]{0.12\linewidth}\centering
		\includegraphics[width=\linewidth]{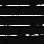}\caption{}\label{res_k}
	\end{subfigure}
	\begin{subfigure}[b]{0.12\linewidth}\centering
		\includegraphics[width=\linewidth]{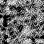}\caption{}\label{res_l}
	\end{subfigure}
	\begin{subfigure}[b]{0.12\linewidth}\centering
		\includegraphics[width=\linewidth]{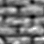}\caption{}\label{res_m}
	\end{subfigure}
	\begin{subfigure}[b]{0.12\linewidth}\centering
		\includegraphics[width=\linewidth]{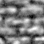}\caption{}\label{res_n}
	\end{subfigure}
\caption{Visualization of inputs and convolutional layers responses. (a, h) and (b, i) are reference patches and distorted patches, respectively. (c, j) are corresponding residual patches. (d, k) and (e, l) are responses of the last 2D convolutional layer correspond to distorted patches and residual patches, respectively. (f, m) are responses of the last 3D convolutional layer. (g, n) are masked residual using learned distortion visibility threshold.} \label{fig:response_visualization}
\end{figure*}

\subsection{Masking and Regression}
\label{ssec:Masking and Regression}

Video quality is primarily decided by quality degradation due to compression artifacts and spatiotemporal masking effects. Psychological study indicates that the HVS cannot perceive small pixel variation in complex backgrounds until the difference reaches a certain level. This is the so-called just-noticeable-difference (JND) threshold. In recent work, the video distortion threshold was learned with hand-crafted features in \cite{wang2018prediction} and 2D CNN in \cite{kim2017deep}. Inspired by the study in human vision \cite{daly1992visible}, the HVS pays more attention to regions with more noticeable distortions. Thus, we mask residual frames with the learned distortion threshold. Accordingly, noticeable distortions would be pronounced only when the masking effect of the background is weak.

The global average pooling layer is used after the distortion threshold masking to represent the degree of perceived distortions. Two fully connected layers are used to learn the nonlinear relationship between perceived distortions and subjective quality. Then, the objective function of the proposed architecture is defined as
\begin{equation}
L(\mathbf{x_n}, y_n; \theta) = \lambda_1 \left || f_{\theta}(\mathbf{x_n}) - y_n \right ||_{2}^{2} + \lambda_2 L_2
\end{equation}
where $\lambda_1$ and $\lambda_2$ are hyper-parameters, $\mathbf{x_n}$ represents the distorted video, $y_n$ is the subjective quality score, $f_\theta(\cdot)$ indicates the prediction system with parameters $\theta$ and $L_2$ indicates regularization term, respectively.

\subsection{Implementation Details}
\label{ssec:implementation}
We use Pytorch to implement the proposed C3DVQA network and train the network from scratch. Training segments are randomly cropped from videos for data augmentation. We select a random temporal position and sample a clip with 60 frames. We apply non-overlapping sliding window to spatially sample each segment. The window size is 112x112 pixels and all sampled segments have the same subjective score with the distorted video. The video is converted to YCbCr and only the Y channel is used for training and validation. Thus, the dimension of a segment is 1 channel $\times$ 60 frames $\times$ 112 pixels $\times$ 112 pixels. We keep the original picture unscaled and avoid applying any other data augmentation techniques that may introduce extra distortions. For testing, the same sampling procedure is applied and the system predicts the quality score for each segment. Then the scores are averaged to get the score of the entire video.

\section{Experimental Results}
\label{sec:results}
We evaluate the proposed method on the commonly used VQA datasets: the LIVE \cite{seshadrinathan2010study} video dataset and the CSIQ \cite{vu2014vis3} video dataset. The Spearman Rank Order Correlation Coefficient (SROCC) and Pearson Linear Correlation Coefficient (PLCC) are used as evaluation criteria. We compare C3DVQA with several state-of-the-art methods and discuss the effectiveness of 3D convolutional layers by replacing C3D with 2D convolutional layers. We also study the effect of the number of sampled frames per segment.

\subsection{Datasets and Training Details}
\label{dataset}
The LIVE \cite{seshadrinathan2010study} dataset consists of 10 reference videos and 150 distorted videos with four types of distortions: wireless distortions, IP distortions, MJPEG-2 compression and H.264 compression.
The CSIQ \cite{vu2014vis3} dataset contains 12 reference videos and 216 distorted videos generated from 6 distortion types: H.264/AVC compression, H.264 video with packet loss rate, MJPEG compression, Wavelet compression, White noise and HEVC compression.

The Adam optimizer \cite{kingma2014adam} is used to back-propagate gradients. The initial learning rate is 1e-4 for the LIVE datasets and 3e-4 for the CSIQ datasets, respectively. The learning rate is multiplied by 0.9 if the loss saturates for 5 epochs. Training is done for 250 epochs and the model with the smallest training loss is used for validation.

We follow a similar procedure as \cite{kim2018deep} to randomly select $80\%$ of the reference videos for training and the remaining $20\%$ is used for validation. Once a reference video is split into the training or testing set, all distorted videos generated from it would be put into the same set. We conduct experiments in a non-distortion-specific manner. The system does not know the distortion types of training videos. We repeat the experiments for 10 times and use the median values of PLCC and SROCC for performance comparison. As recommended in \cite{series2012methodology}, a non-linear logistic regression function is applied on the predicted score.

%%%%%%%%%%%%%%%%%%%%%%%%%%%%%%%%%%%%%%%%%%%%%%%%%%%%%%%%%%%%%%%%%%
\begin{figure}[!thb]
	\centering
		\begin{subfigure}[b]{1.0\linewidth}
		\centering{}
		\includegraphics[width=0.92\linewidth]{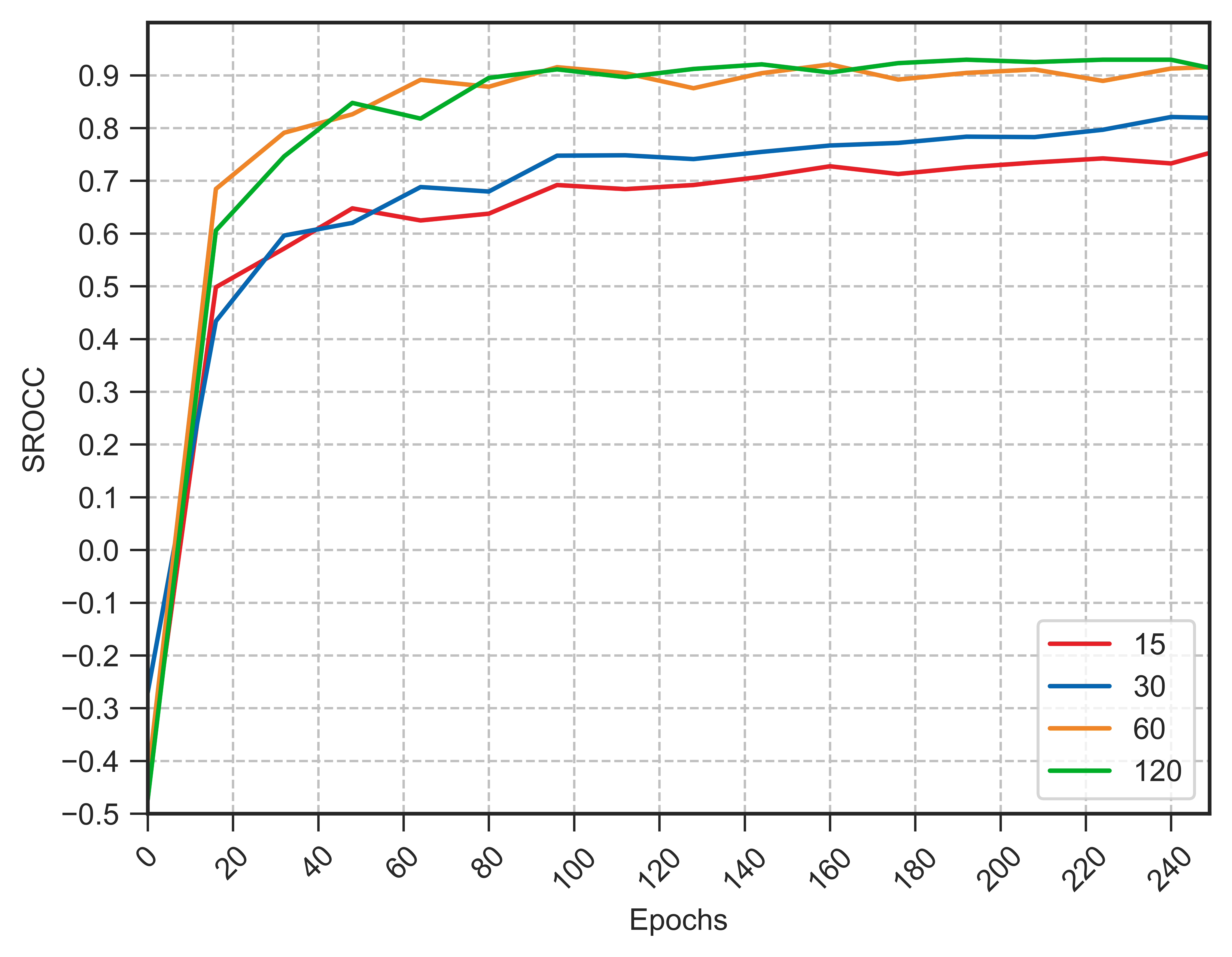}
		% \caption*{\label{}}
		\end{subfigure}
	\caption{SROCC curves when C3DVQA is trained with the following number of frames: 15, 30, 60, 120.} \label{fig:frame_number}
\end{figure}
%%%%%%%%%%%%%%%%%%%%%%%%%%%%%%%%%%%%%%%%%%%%%%%%%%%%%%%%%%%%%%%%%%

We visualize layer responses of the system in Fig. \ref{fig:response_visualization}, where (f, m) and (g, n) are learned spatiotemporal threshold and the masked residual patches, respectively. Brighter pixels indicate distortions that are more noticeable to the HVS. We can observe that the hair area of the girl is almost black in Fig. \ref{fig:response_visualization} (g) and distortions are hardly visible. The artifacts in the central area of the brick are noticeable and the bright regions in Fig. \ref{fig:response_visualization} (n) overlaps with those areas.

\subsection{Comparison with the State-of-the-Art }
\label{dataset}
We compare the performance of C3DVQA against a number of FR-VQA models: PSNR, MOVIE \cite{seshadrinathan2009motion}, ST-MAD \cite{vu2011spatiotemporal}, VMAF \cite{li2016toward}, and DeepVQA \cite{kim2018deep}. The results are given in Table \ref{tab:result_live_csiq}. The model trained with the C3D achieves the better or comparable results on both datasets. DeepVQA is a FR-VQA metric using CNN and the attention mechanism. It is clear that CNN based methods outperform feature-engineering-based methods by a large margin. PSNR is a signal-integrity-based method and ignores the motion information of videos. MOVIE and ST-MAD explicitly exploit motion information and outperform IQA based methods. VMAF is built on top of several IQA metrics and aggregates frame scores to get the score of the whole video. It adopts a frame difference between adjacent frames to account for the motion information. There is no surprise that it performs better than IQA metrics.

\subsection{Ablation Study}
\label{ablation experiments}
We validate the effectiveness of 3D convolutional layers. We replace 3D convolutional layers with 2D convolutional layers and average frame scores to get the video score. Evaluation results are indicated by C3DVQA (2D) in Table \ref{tab:result_live_csiq}. Without exploiting motion information between adjacent frames, C3DVQA (2D) is effectively an IQA metric that learns to predict frame quality scores. As expected, the performance drops slightly. We also find the same phenomenon while reproducing results of the DeepVQA method. The temporal module boots the performance by $2\%$ in terms of PLCC and SROCC.

We empirically find that the number of sampled frames has a great impact on the performance and computational costs. We repeated the experiments with the following number of frames: 15, 30, 60 and 120 and Fig. \ref{fig:frame_number} depicts SROCC of each setting over 250 epochs. The best performance is obtained when the segment length is 60 or higher.
%However, As shown in Table \ref{tab:frames_compare}, the training time apparently increases when the segment has more frames.

%%%%%%%%%%%%%%%%%%%%%%%%%%%%%%%%%%%%%%%%%%%%%%%%%%%%%%%%%%%%%%
\begin{table}
	\begin{center}
	\caption{
	Effects of segment length for videos in the CSIQ dataset. The last row indicates the training time (seconds) for 1 epoch. \label{tab:frames_compare}
	}
		\begin{tabular}{|c|c|c|c|c|}
		\hline
                & 15        & 30        & 60        & 120\\
		\hline
		PLCC    & 0.7220    & 0.8279    & 0.9043    & 0.9230\\
		\hline
		SROCC   & 0.7423    & 0.8360    & 0.9152    & 0.9331\\
		\hline
		Time (sec)    & 199    & 202     & 350    & 412 \\
		\hline
		\end{tabular}
	\end{center}
\end{table}
%%%%%%%%%%%%%%%%%%%%%%%%%%%%%%%%%%%%%%%%%%%%%%%%%%%%%%%%%%%%%%%%%%
%%%%%%%%%%%%%%%%%%%%%%%%%%%%%%%%%%%%%%%%%%%%%%%%%%%%%%%%%%%%%%%%%%
\begin{table}
	\begin{center}
		\caption{
		Performance comparison (median) on the LIVE and CSIQ video datasets. The top performing algorithm are highlighted in bold font.\label{tab:result_live_csiq}
		}
		\begin{tabular}{|c|c|c|c|c|}
			\hline
			\multicolumn{1}{|c|}{} &
			\multicolumn{2}{|c|}{LIVE} &
			\multicolumn{2}{|c|}{CSIQ} \\
			\cline{2-3}
			\cline{4-5}
			Methods & PLCC & SROCC & PLCC & SROCC\\
			\hline
			PSNR    & 0.7271   & 0.7398   & 0.5988   & 0.6106\\
			\hline
			MOVIE \cite{seshadrinathan2009motion}  & 0.8609    & 0.8481    & 0.6295    & 0.6247\\
			ST-MAD \cite{vu2011spatiotemporal} & 0.8570 & 0.8386 & 0.7674  & 0.7766\\
			\hline
			VMAF \cite{li2016toward}   & 0.8115 & 0.8163 & 0.6570 & 0.6377\\
			\hline
			DeepVQA \cite{kim2018deep} & 0.8952 & 0.9152 & \textbf{0.9135} & 0.9123\\
			C3DVQA (2D)  & 0.8674 & 0.8885 & 0.8554 & 0.8879\\
			C3DVQA (3D)  & \textbf{0.9122} & \textbf{0.9261} & 0.9043 & \textbf{0.9152}\\
			\hline
		\end{tabular}
	\end{center}
\end{table}
% \vspace{-4em}

\section{Conclusion and Future Work}
\label{sec:conclusion}
We propose an end-to-end spatiotemporal feature learning framework using 3D CNN for the full-reference VQA task. We use 2D convolutional layers to learn spatial features and 3D convolutional layers to learn the spatiotemporal distortion threshold. Residual frames are masked with the distortion threshold to mimic the quality evaluation process of the HVS. Finally, we use nonlinear regression to map the predicted score to the subjective score. Experimental results are given to demonstrate the performance of the proposed architecture.
We also show the effectiveness of 3D convolutional layers and study the performance of C3DVQA with different segment length in ablation study.
In the future, we plan to generalize the proposed network to the large-scale VQA dataset \cite{wang2017videoset} and conduct more comprehensive benchmarking experiments. We are also interested in interpreting the learned features with respect to different distortions types.

\section{Acknowledgement}
This research received support from the Shenzhen Municipal Development and Reform Commission (Disciplinary Development Program for Data Science and Intelligent Computing), Shenzhen Municipal Science and Technology Program (No. JCYJ20170818141146428) and National Engineering Laboratory for Video Technology - Shenzhen Division.

\newpage
\clearpage

\bibliographystyle{IEEEbib}
\bibliography{refs}

\end{document}